\documentclass[aps,preprint,showpacs,showkeys,amsmath,amssymb,nofootinbib,byrevtex]{revtex4}
\usepackage{graphicx,subfigure}
\DeclareGraphicsExtensions{.eps}

\begin{document}


\title{$\kappa$-GENERALIZED STATISTICS IN PERSONAL INCOME DISTRIBUTION}
\author{F. Clementi}
\email[\textit{Corresponding author}: ]{fabio.clementi@univpm.it}
\author{M. Gallegati}
\affiliation{Department of Economics, Universit\`a Politecnica delle Marche, Piazzale Martelli 8, 60121 Ancona, Italy}
\author{G. Kaniadakis}
\affiliation{Department of Physics, Politecnico di Torino, Corso Duca degli Abruzzi 24, 10129 Torino, Italy}
\date{\today}


\begin{abstract}
Starting from the generalized exponential function $\exp_{\kappa}\left(x\right)=\left(\sqrt{1+\kappa^{2}x^{2}}+\kappa x\right)^{1/\kappa}$, with $\exp_{0}\left(x\right)=\exp(x)$, proposed in Ref. [G. Kaniadakis, Physica A \textbf{296}, 405 (2001)], the survival function $P_{>}\left(x\right)=\exp_{\kappa}\left(-\beta x^{\alpha}\right)$, where $x\in\mathbf{R}^{+}$, $\alpha,\beta>0$, and $\kappa\in\left[0,1\right)$, is considered in order to analyze the data on personal income distribution for Germany, Italy, and the United Kingdom. The above defined distribution is a continuous one-parameter deformation of the stretched exponential function $P_{>}^{0}\left(x\right)=\exp\left(-\beta x^{\alpha}\right)$\textemdash to which reduces as $\kappa$ approaches zero\textemdash behaving in very different way in the $x\rightarrow0$ and $x\rightarrow\infty$ regions. Its bulk is very close to the stretched exponential one, whereas its tail decays following the power-law $P_{>}\left(x\right)\sim\left(2\beta\kappa\right)^{-1/\kappa}x^{-\alpha/\kappa}$. This makes the $\kappa$-generalized function particularly suitable to describe simultaneously the income distribution among both the richest part and the vast majority of the population, generally fitting different curves. An excellent agreement is found between our theoretical model and the observational data on personal income over their entire range.
\end{abstract}
\maketitle


\section{Introduction}
A renewed interest in studying the distribution of income has emerged over the last years in both the physics and economics communities \cite{ChatterjeeYarlagaddaChakraborti2005}. The focus has been mostly put on empirical analysis of extensive datasets to infer the exact shape of personal income distributions, and to design theoretical models that can reproduce them \cite{RichmondHutzlerCoelhoRepetowicz2006}.
\par
A natural starting point in this area of enquiry was the observation that the number of persons in a population whose incomes exceed $x$ is often well approximated by $Cx^{-\alpha}$, for some real $C$ and some positive $\alpha$, as Pareto \cite{Pareto1896,Pareto1897a,Pareto1897b} argued over 100 years ago. However, theoretical and empirical work rapidly pointed out the fact that it is only in the upper tail of the income distribution that a Pareto-like behavior can be expected \cite{Arnold1983}, while the bulk of the income\textemdash held by the 95\% or so of the population\textemdash is governed by a completely different law. Therefore, many recent papers within this literature have sought to characterize the distribution of income by a mixture of known statistical distributions, even if there is a dispute about what these distributions are: indeed, while it seems to be generally acknowledged that the top 1--5\% of incomes follows the Pareto's law, an exact and unequivocal characterization of the low to medium income region of the distribution is still evasive. For example, Refs. \cite{ClementiGallegati2005a,ClementiGallegati2005b,AitchisonBrown1954,AitchisonBrown1957,DiMatteoAsteHyde2004,Gibrat1931,MontrollShlesinger1982,MontrollShlesinger1983,Souma2001} claim that this is lognormal, while according to Refs. \cite{DragulescuYakovenko2001a,DragulescuYakovenko2001b,NireiSouma2004,SilvaYakovenko2005,WillisMimkes2004} the distribution of personal income for the majority of the population should follow the exponential law.
\par
In the present work we address the issue of data analysis related to the size distribution of income by adopting a statistical mechanics approach
introduced by one of us in Refs. \cite{Kaniadakis2001a,Kaniadakis2001b,Kaniadakis2002,Kaniadakis2005,Kaniadakis2006}, based on the one-parameter generalization of the exponential function defined through
\begin{equation}
\exp_{\kappa}\left(x\right)=\left(\sqrt{1+\kappa^{2} x^{2}}+\kappa x\right)^{1/\kappa},\quad x\in\mathbf{R}.
\label{eq:Equation1}
\end{equation}
The properties of the function $\exp_{\kappa}\left(x\right)\in C^{\infty}\left(\mathbf{R}\right)$ have been considered extensively in the literature. We recall briefly that in the $\kappa\rightarrow0$ limit the function $\exp_{\kappa}\left(x\right)$ reduces to the ordinary exponential, \textit{i.e.} $\exp_{0}\left(x\right)=\exp\left(x\right)$, and for $x\rightarrow0$\textemdash independently on the value of $\kappa$\textemdash behaves very similarly with the ordinary exponential, holding for $\kappa^{2}x^{2}<1$ the following Taylor expansion
\begin{equation}
\exp_{\kappa}\left(x\right)=1+x+\frac{x^{2}}{2}+\left(1-\kappa^{2}\right)\frac{x^{3}}{3!}+\ldots\quad.
\label{eq:Equation2}
\end{equation}
It is remarkable that the first three terms of the Taylor expansion are the same as the ordinary exponential. On the other hand, the most interesting property of $\exp_{\kappa}\left(x\right)$ for the applications in statistics is the power-law asymptotic behavior
\begin{equation}
\exp_{\kappa}\left(x\right){\atop\stackrel{\textstyle\sim}{\scriptstyle x\rightarrow\pm\infty}}\left|2\kappa x\right|^{\pm1/\left|\kappa\right|}.
\label{eq:Equation3}
\end{equation}
\par
The generalized logarithmic function $\ln_{\kappa}\left(x\right)\in C^{\infty}\left(\mathbf{R}^{+}\right)$ is defined as the inverse function of $\exp_{\kappa}\left(x\right)$, namely $\ln_{\kappa}\left(\exp_{\kappa}x\right)=\exp_{\kappa}\left(\ln_{\kappa}x\right)=x$, and is given by
\begin{equation}
\ln_{\kappa}\left(x\right)=\frac{x^{\kappa}-x^{-\kappa}}{2\kappa}.
\label{eq:Equation4}
\end{equation}
\par
Starting from the generalized logarithm, the new entropy
\begin{equation}
S\left(f\right)=-\left\langle \ln_{\kappa}\left(f\right)\right\rangle=-\int\mathrm{d}xf\left(x\right)\ln_{\kappa}\left(f(x\right)
\label{eq:Equation5}
\end{equation}
has been introduced, which can be written explicitly as
\begin{equation}
S_{\kappa}=\int\mathrm{d}x\frac{f\left(x\right)^{1-\kappa}-f\left(x\right)^{1+\kappa}}{2\kappa},
\label{eq:Equation6}
\end{equation}
being $f\left(x\right)$ the probability distribution function. The latter entropy has the standard properties of the ordinary Boltzmann-Shannon entropy (which recovers in the $\kappa\rightarrow0$ limit): is thermodynamically-stable, is Lesche-stable, obeys the Khinchin axioms of continuity, maximality, expandability and generalized additivity.
\par
After maximizing the entropy \eqref{eq:Equation6} under the proper constraints according to the Jaynes Maximum Entropy Principle of statistical mechanics, the probability distribution function
\begin{equation}
p\left(x\right)=\alpha\exp_{\kappa}\left(-\frac{E\left(x\right)-\mu}{\lambda k_{B}T}\right)
\label{eq:Equation7}
\end{equation}
obtains, where
\begin{equation}
\lambda=\sqrt{1-\kappa^{2}}\quad\text{and}\quad\alpha=\left(\frac{1-\kappa}{1+\kappa}\right)^{1/2\kappa}.
\label{eq:Equation8}
\end{equation}
For a particle system, $x$ represents the particle velocity, $E\left(x\right)$ the energy, $\mu$ the chemical potential, $T$ the temperature, and $k_{B}$ the Boltzmann constant.
\par
Also the distribution function
\begin{equation}
f\left(x\right)=\frac{1}{Z}\exp_{\kappa}\left(-\beta x^{\alpha}\right)
\label{eq:Equation9}
\end{equation}
has been considered to define both probability distribution functions (with $Z$ a normalization constant) as well as cumulative distribution functions (with $Z=1$). The distribution functions given by Eqs. \eqref{eq:Equation7} and \eqref{eq:Equation9} have been used to analyze also non-physical systems.
\par
The main result of the present effort is that the cumulative distribution function defined by Eq. \eqref{eq:Equation9} can describe the whole spectrum of the size distribution of income, ranging from the low region to the middle region, and up to the power-law tail, pointing in this way toward a unified approach to the problem.
\par
The paper is organized as follows. In Sec. \ref{sec:SectionII} we consider the main properties of the $\kappa$-generalized statistical distribution functions. In Sec. \ref{sec:SectionIII}, in order to asses the reliability of the proposed $\kappa$-distribution, we compare the theoretical curve with the census data for personal income in Germany, Italy, and the United Kingdom. Finally, in Sec. \ref{sec:SectionIV} some concluding remarks are reported.


\setcounter{equation}{0}
\section{The $\kappa$-generalized statistical distribution}
\label{sec:SectionII}
\begin{figure}[!t]
\centering
\begin{center}
\mbox{
\subfigure[\hspace{0.2cm}CCDF for some different values of $\beta$]{\label{fig:Figure1_a}\includegraphics[width=0.48\textwidth]{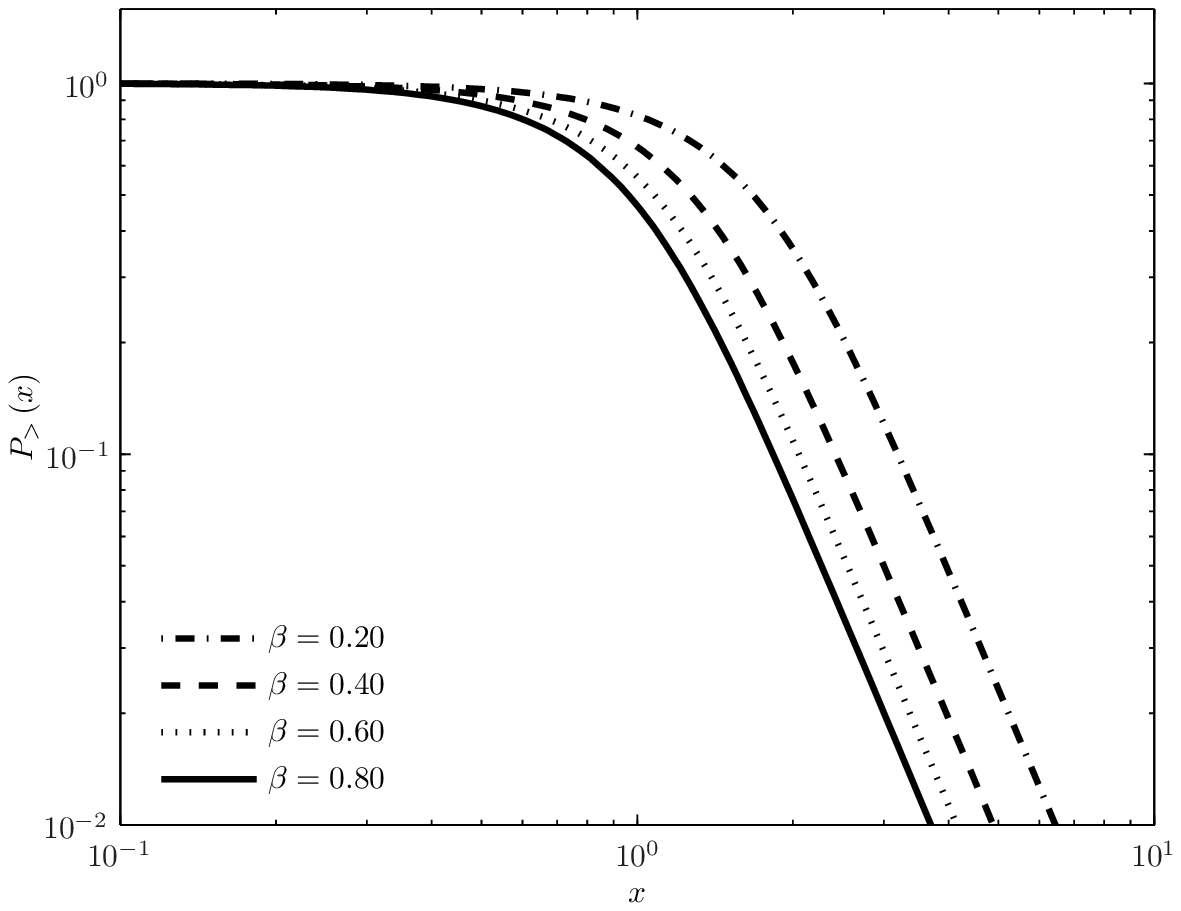}}
\subfigure[\hspace{0.2cm}PDF for some different values of $\beta$]{\label{fig:Figure1_b}\includegraphics[width=0.48\textwidth]{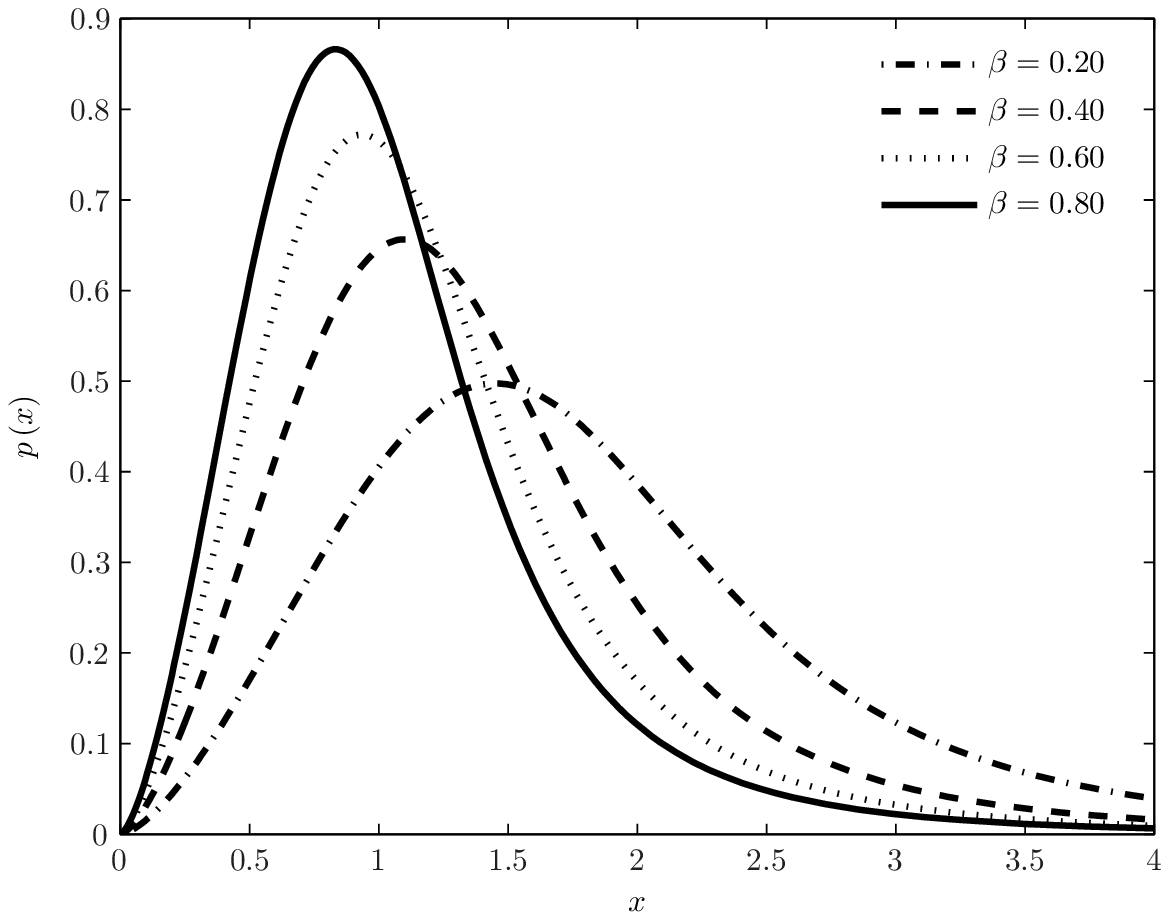}}
}
\caption{\subref{fig:Figure1_a} Plot of the $\kappa$-generalized CCDF given by Eq. \eqref{eq:Equation10} versus $x$ for some different values of $\beta$ ($=0.20,0.40,0.60,0.80$), and fixed $\alpha$ ($=2.50$) and $\kappa$ ($=0.75$). \subref{fig:Figure1_b} Plot of the $\kappa$-generalized PDF given by Eq. \eqref{eq:Equation13} versus $x$ for some different values of $\beta$ ($=0.20,0.40,0.60,0.80$), and fixed $\alpha$ ($=2.50$) and $\kappa$ ($=0.75$). Notice that the distribution spreads out (concentrates) as the value of $\beta$ decreases (increases)}
\label{fig:Figure1}
\end{center}
\end{figure}

\begin{figure}[t]
\centering
\begin{center}
\mbox{
\subfigure[\hspace{0.2cm}CCDF for some different values of $\alpha$]{\label{fig:Figure2_a}\includegraphics[width=0.48\textwidth]{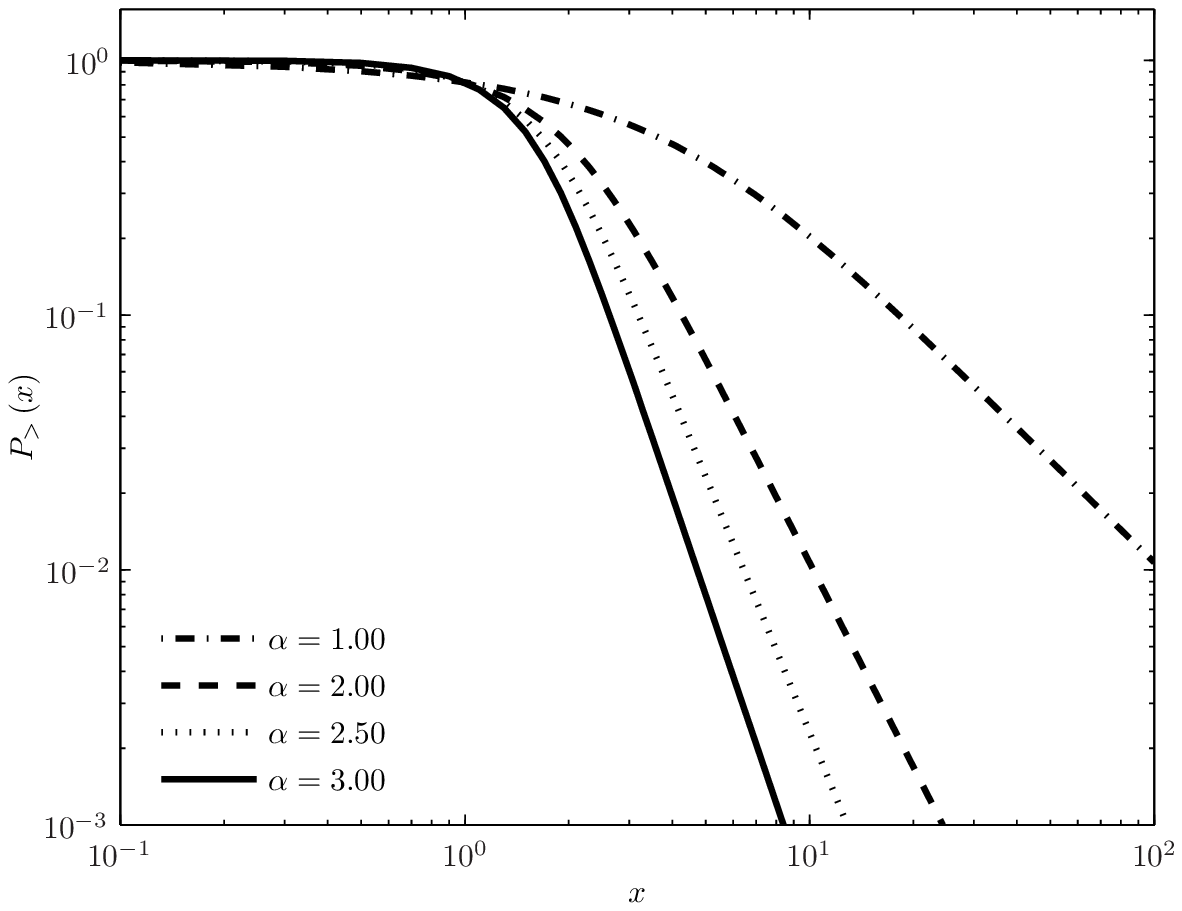}}
\subfigure[\hspace{0.2cm}PDF for some different values of $\alpha$]{\label{fig:Figure2_b}\includegraphics[width=0.48\textwidth]{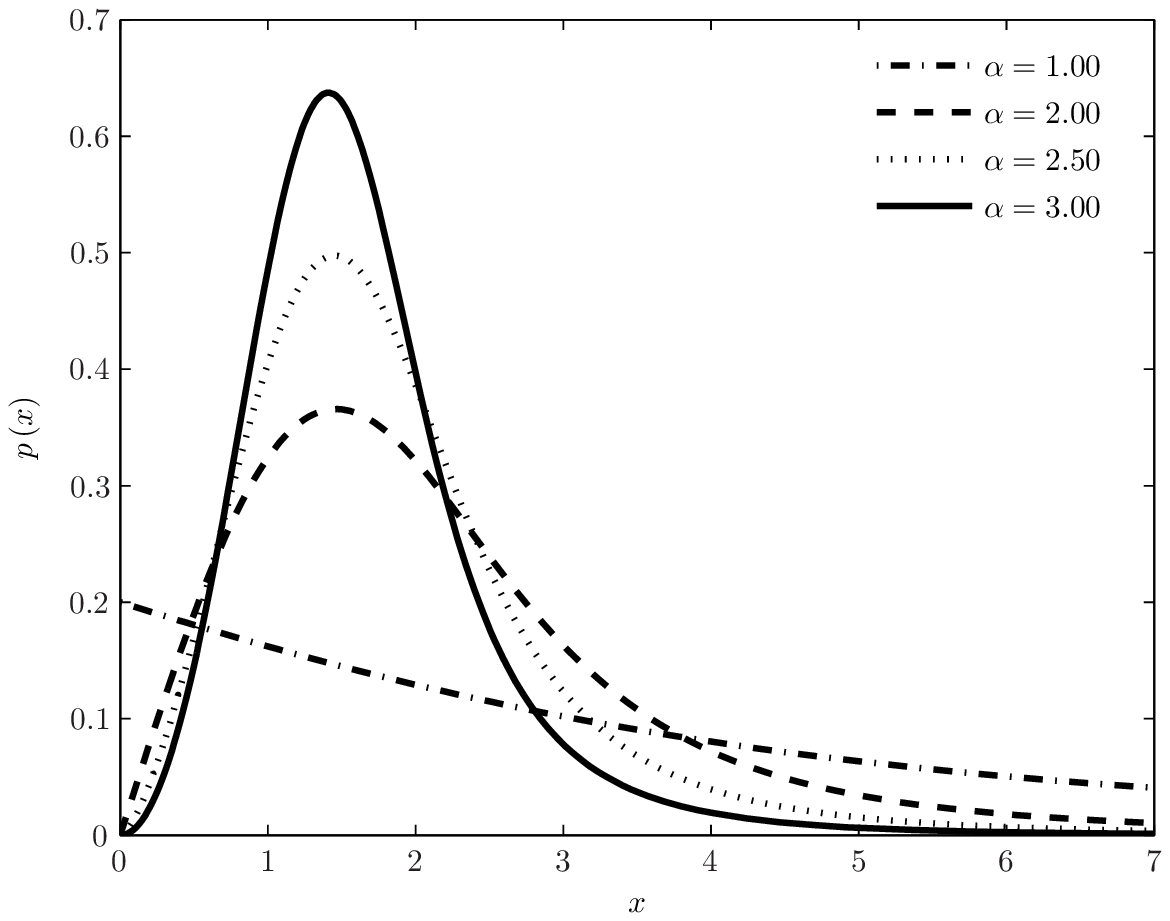}}
}
\caption{\subref{fig:Figure2_a} Plot of the $\kappa$-generalized CCDF given by Eq. \eqref{eq:Equation10} versus $x$ for some different values of $\alpha$ ($=1.00,2.00,2.50,3.00$), and fixed $\beta$ ($=0.20$) and $\kappa$ ($=0.75$). \subref{fig:Figure2_b} Plot of the $\kappa$-generalized PDF given by Eq. \eqref{eq:Equation13} versus $x$ for some different values of $\alpha$ ($=1.00,2.00,2.50,3.00$), and fixed $\beta$ ($=0.20$) and $\kappa$ ($=0.75$). Notice that the curvature (shape) of the distribution becomes less (more) pronounced when the value of $\alpha$ decreases (increases). The case $\alpha=1.00$ corresponds to the ordinary exponential function}
\label{fig:Figure2}
\end{center}
\end{figure}

\begin{figure}[t]
\centering
\begin{center}
\mbox{
\subfigure[\hspace{0.2cm}CCDF for some different values of $\kappa$]{\label{fig:Figure3_a}\includegraphics[width=0.48\textwidth]{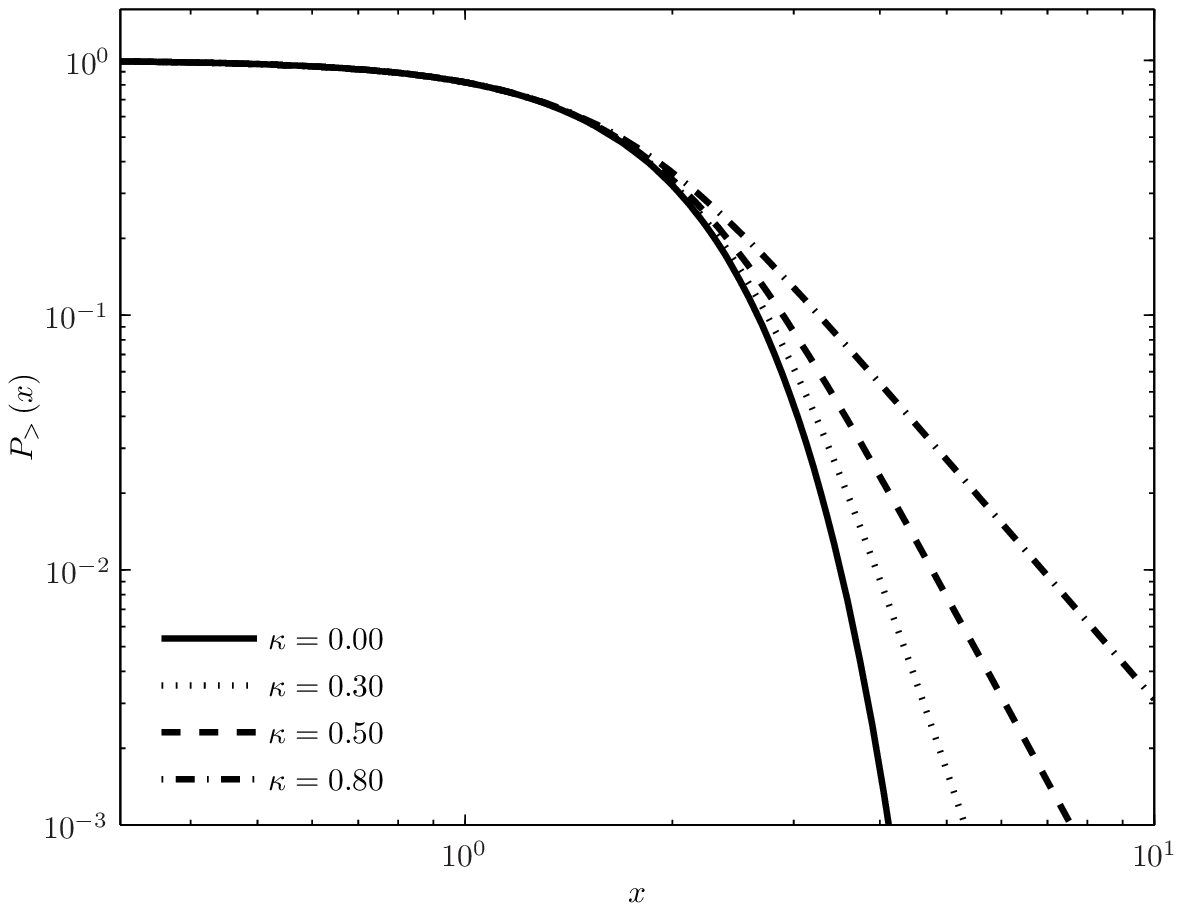}}
\subfigure[\hspace{0.2cm}PDF for some different values of $\kappa$]{\label{fig:Figure3_b}\includegraphics[width=0.48\textwidth]{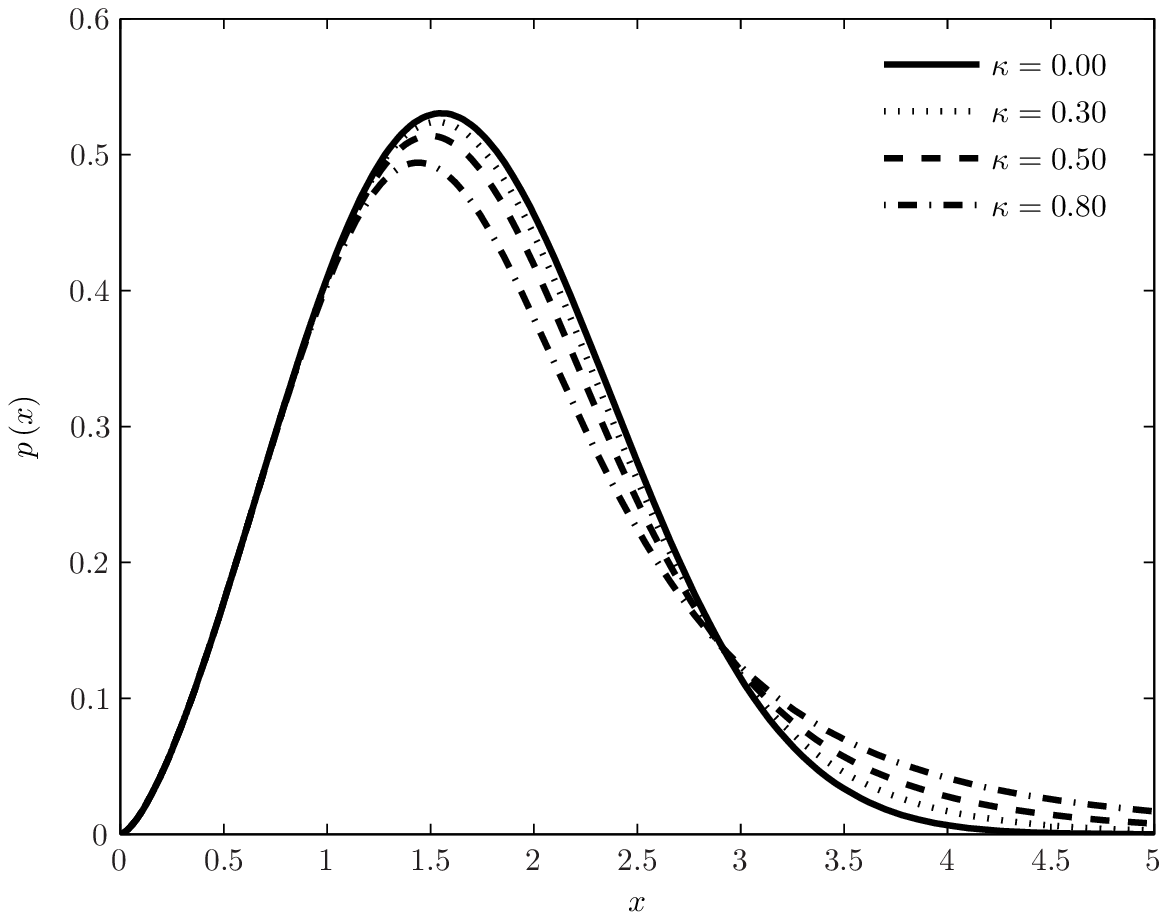}}
}
\caption{\subref{fig:Figure3_a} Plot of the $\kappa$-generalized CCDF given by Eq. \eqref{eq:Equation10} versus $x$ for some different values of $\kappa$ ($=0.00,0.30,0.50,0.80$), and fixed $\beta$ ($=0.20$) and $\alpha$ ($=2.50$). \subref{fig:Figure3_b} Plot of the $\kappa$-generalized PDF given by Eq. \eqref{eq:Equation13} versus $x$ for some different values of $\kappa$ ($=0.00,0.30,0.50,0.80$), and fixed $\beta$ ($=0.20$) and $\alpha$ ($=2.50$). Notice that the upper tail of the distribution fattens (thins) as the value of $\kappa$ increases (decreases). The case $\kappa=0.00$ corresponds to the ordinary stretched exponential (Weibull) function \cite{LaherrereSornette1998,Sornette2000}}
\label{fig:Figure3}
\end{center}
\end{figure}

The $\kappa$-generalized Complementary Cumulative Distribution Function (CCDF) is given by
\begin{equation}
P_{>}\left(x\right)=\exp_{\kappa}\left(-\beta x^{\alpha}\right),\quad x\in\mathbf{R}^{+},
\label{eq:Equation10}
\end{equation}
being $P_{>}\left(x\right)$ the probability of finding the distribution variable with a value $X$ greater than $x$. The income variable $x$ is defined as $x=z/\left\langle z\right\rangle$, being $z$ the absolute personal income and $\left\langle z\right\rangle$ its mean value. Then the dimensionless variable $x$ represents the personal income in units of $\left\langle z\right\rangle$. The constant $\beta>0$ is a characteristic scale, since its value determines the scale of the probability distribution: if $\beta$ is large, then the distribution will be more concentrated; if $\beta$ is small, then it will be more spread out (see FIG. \ref{fig:Figure1_a}--\subref{fig:Figure1_b}).
The exponent $\alpha>0$ quantifies the curvature (shape) of the distribution, which is less (more) pronounced for lower (higher) values of the parameter, as seen in FIG. \ref{fig:Figure2_a}--\subref{fig:Figure2_b}.
Finally, as one can observe in FIG. \ref{fig:Figure3_a}--\subref{fig:Figure3_b}, the deformation parameter $\kappa\in\left[0,1\right)$ measures the fatness of the upper tail: the larger (smaller) its magnitude, the fatter (thinner) the tail.
\par
The function $P_{>}\left(x\right)$ defined through Eq. \eqref{eq:Equation10} can be viewed as a generalization of the ordinary stretched exponential \cite{LaherrereSornette1998}, \textit{i.e.} $P_{>}^{0}\left(x\right)=\exp\left(-\beta x^{\alpha}\right)$, which recovers in the $\kappa\rightarrow0$ limit. It is remarkable that $P_{>}\left(x\right)$ for $x\rightarrow0^{+}$ behaves as the ordinary stretched exponential
\begin{equation}
P_{>}\left(x\right){\atop\stackrel{\textstyle\sim}{\scriptstyle x\rightarrow{0^{+}}}}\exp\left(-\beta x^{\alpha}\right),
\label{eq:Equation11}
\end{equation}
while for $x\rightarrow\infty$ presents a power-law tail
\begin{equation}
P_{>}\left(x\right){\atop\stackrel{\textstyle\sim}{\scriptstyle x\rightarrow{+\infty}}}\left(2\beta\kappa\right)^{-1/\kappa}x^{-\alpha/\kappa}.
\label{eq:Equation12}
\end{equation}
\par
The Probability Density Function (PDF), $p\left(x\right)=-\mathrm{d}P_{>}\left(x\right)/\mathrm{d}x$, is given by
\begin{equation}
p\left(x\right)=\frac{\alpha\beta x^{\alpha-1}\exp_{\kappa}\left(-\beta x^{\alpha}\right)}{\sqrt{1+\beta^{2}\kappa^{2}x^{2\alpha}}},
\label{eq:Equation13}
\end{equation}
and can viewed as a generalization of the Weibull distribution \cite{Sornette2000}, \textit{i.e.} $p^{0}\left(x\right)=\alpha\beta x^{\alpha-1}\exp\left(-\beta x^{\alpha}\right)$, which recovers in the $\kappa\rightarrow0$ limit. The function $p\left(x\right)$ given by Eq. \eqref{eq:Equation13} for $x\rightarrow0^{+}$ behaves as a Weibull distribution
\begin{equation}
p\left(x\right){\atop\stackrel{\textstyle\sim}{\scriptstyle x\rightarrow{0^{+}}}}\alpha\beta x^{\alpha-1}\exp\left(-\beta x^{\alpha}\right),
\label{eq:Equation14}
\end{equation}
while for $x\rightarrow+\infty$ reduces to the Pareto's law
\begin{equation}
p\left(x\right){\atop\stackrel{\textstyle\sim}{\scriptstyle x\rightarrow{+\infty}}}\frac{\alpha}{\kappa}\left(2\beta\kappa\right)^{-1/\kappa}x^{-\left(\frac{\alpha}{\kappa}+1\right)}.
\label{eq:Equation15}
\end{equation}
\par
Starting from the law \eqref{eq:Equation13}, one can calculate the mean value $\left\langle x\right\rangle$ which, taking into account the meaning of the variable $x$, results to be equal to unity
\begin{equation}
\int\limits_{0}^{\infty}xp\left(x\right)\mathrm{d}x=1.
\label{eq:Equation16}
\end{equation}
The latter relationship permits to express the parameter $\beta$ as a function of the parameters $\kappa$ and $\alpha$, obtaining
\begin{equation}
\beta=\frac{1}{2\left|\kappa\right|}\left[\frac{\Gamma\left(\frac{1}{\alpha}\right)}{\left|\kappa\right|+\alpha}\frac{\Gamma\left(\frac{1}{2|\kappa|}-\frac{1}{2\alpha}\right)}{\Gamma\left(\frac{1}{2|\kappa|}+\frac{1}{2\alpha}\right)}\right]^{\alpha},
\label{eq:Equation17}
\end{equation}
where $\Gamma\left(x\right)$ is the Euler gamma function $\Gamma\left(x\right)=\int_{0}^{\infty}t^{x-1}e^{-t}\mathrm{d}t$. Thus the problem to determine the values of the free parameters ($\kappa$, $\alpha$, $\beta$) of the theory from the empirical data reduces to a two parameter ($\kappa$, $\alpha$) fitting problem.


\section{An application to personal income data}
\label{sec:SectionIII}
As a working example, we analyze the census data on personal income in three countries: Germany, Italy, and the United Kingdom.\footnote{See Refs. \cite{ClementiGallegati2005a,ClementiGallegati2005b,ClementiDiMatteoGallegati2006} for analysis referring to the same countries and data sources.}
\par
The data used are drawn primarily from the Cross-National Equivalent File (CNEF) 1980--2002, a commercially available dataset compiled by researchers at Cornell University which attempts to make comparable, among others, the following panel surveys: the German Socio-Economic Panel (GSOEP) and the British Household Panel Study (BHPS).\footnote{For background on the CNEF, see Ref. \cite{BurkhauserButricaDalyLillard2000} or consult the CNEF homepage at the following web address: \url{http://www.human.cornell.edu/che/PAM/Research/Centers-Programs/German-Panel/cnef.cfm}.} The income variable we use is the post-government income, representing the combined income after taxes and government transfers of the head, partner, and
other family members.
\par
For Italy, which is not part of the CNEF, we use the Survey on Household Income and Wealth (SHIW), a household-based panel study carried out by the Bank of Italy since 1977. In place of the post-government income in the CNEF, we use the net disposable income variable from the survey above\textemdash \textit{i.e.}, the income recorded after the payment of taxes and social security contributions, defined as the sum of four main components: compensation of employees, pensions and net transfers, net income from self-employment, and property income.\footnote{For a comprehensive discussion of the dataset, see Ref. \cite{Brandolini1999}; the data are available for free download at the following web address: \url{http://www.bancaditalia.it/statistiche/ibf/statistiche/ibf/microdati/dati/en_archivio.htm}.} 
\par
\begin{figure}[t]
\centering
\begin{center}
\mbox{
\subfigure[\hspace{0.2cm}Complementary CDF]{\label{fig:Figure4_a}\includegraphics[width=0.48\textwidth]{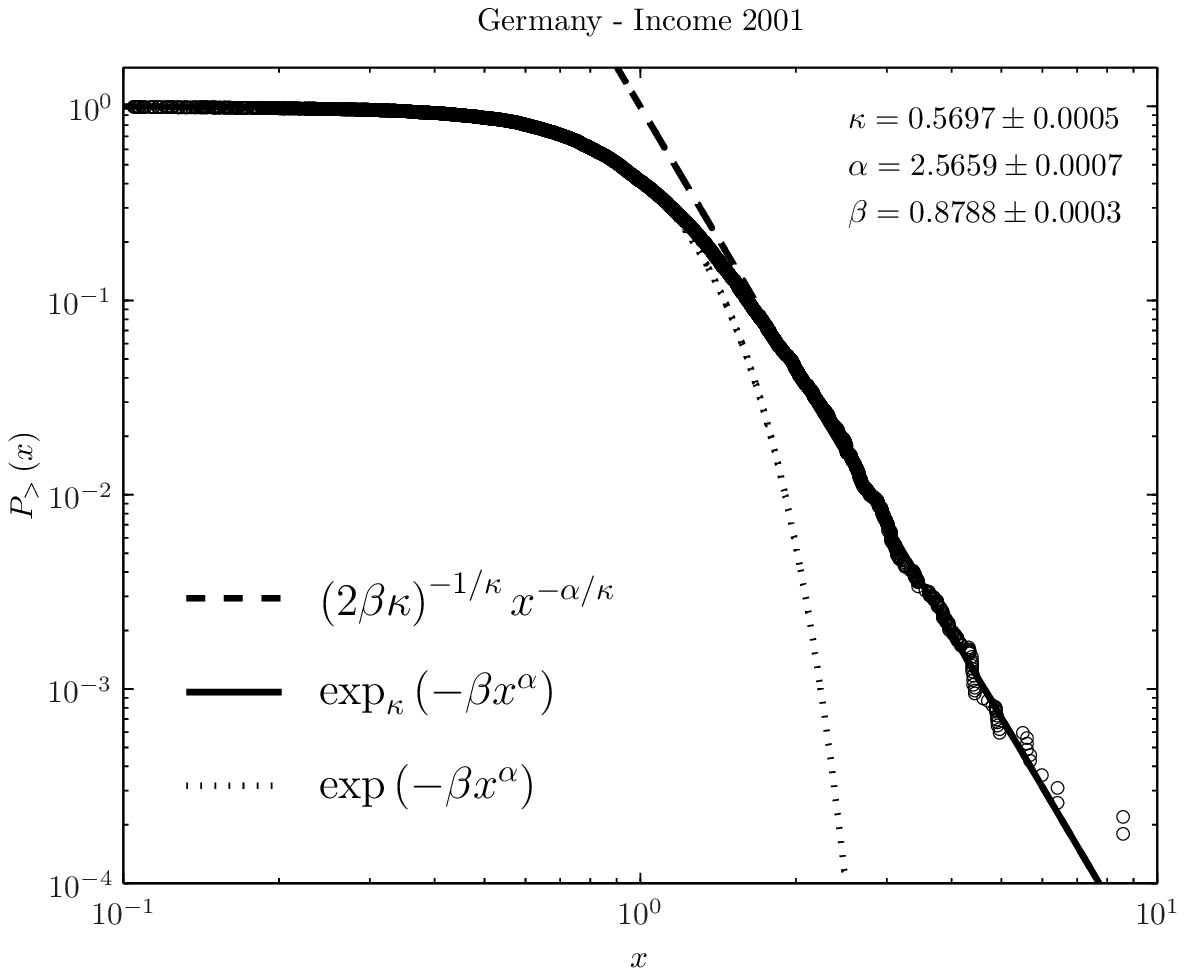}}
\subfigure[\hspace{0.2cm}PDF histogram plot]{\label{fig:Figure4_b}\includegraphics[width=0.48\textwidth]{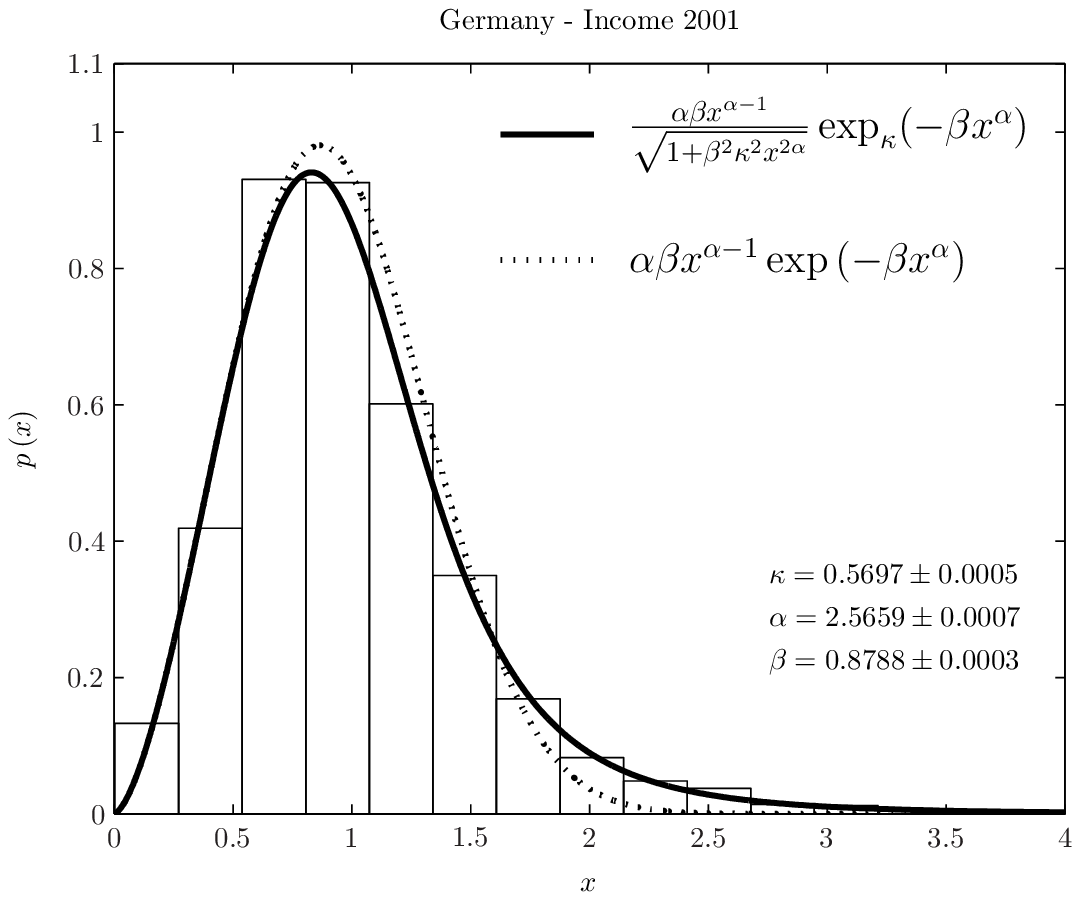}}}
\caption{The German personal income distribution from the 2001 GSOEP-CNEF data file measured in current year euros. \subref{fig:Figure4_a} Plot of the empirical CCDF versus income in the log-log scale. The solid line is our theoretical model given by Eq. \eqref{eq:Equation10} with $\kappa=0.5697\pm0.0005$, $\alpha=2.5659\pm0.0007$, and $\beta=0.8788\pm0.0003$, which fits very well the data in the whole range from the low to the high incomes including the intermediate income region. This function is compared with the ordinary stretched exponential one (dotted line)\textemdash fitting the low income data\textemdash and with the pure power-law (dashed line)\textemdash fitting the high income data\textemdash by using the same parameter values. \subref{fig:Figure4_b} Histogram plot of the empirical PDF with superimposed fits of the $\kappa$-generalized (solid line) and stretched exponential (dotted line) PDFs using the same parameter values as in panel \subref{fig:Figure4_a}. The income axis limits have been adjusted according to the range of data to shed light on the intermediate region between the bulk and the tail of the distribution}
\label{fig:Figure4}
\end{center}
\end{figure}
\begin{figure}[t]
\centering
\begin{center}
\mbox{
\subfigure[\hspace{0.2cm}Complementary CDF]{\label{fig:Figure5_a}\includegraphics[width=0.48\textwidth]{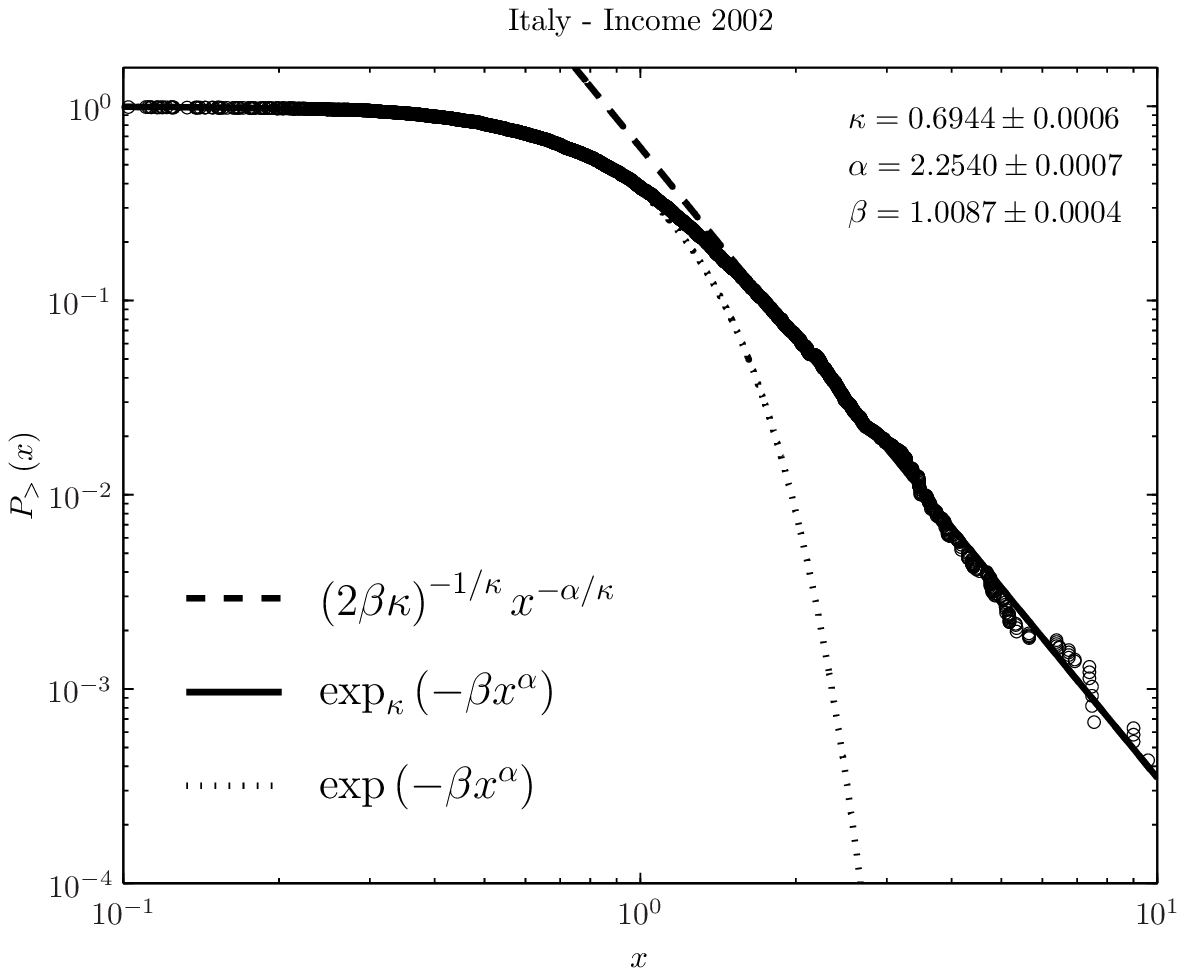}}
\subfigure[\hspace{0.2cm}PDF histogram plot]{\label{fig:Figure5_b}\includegraphics[width=0.48\textwidth]{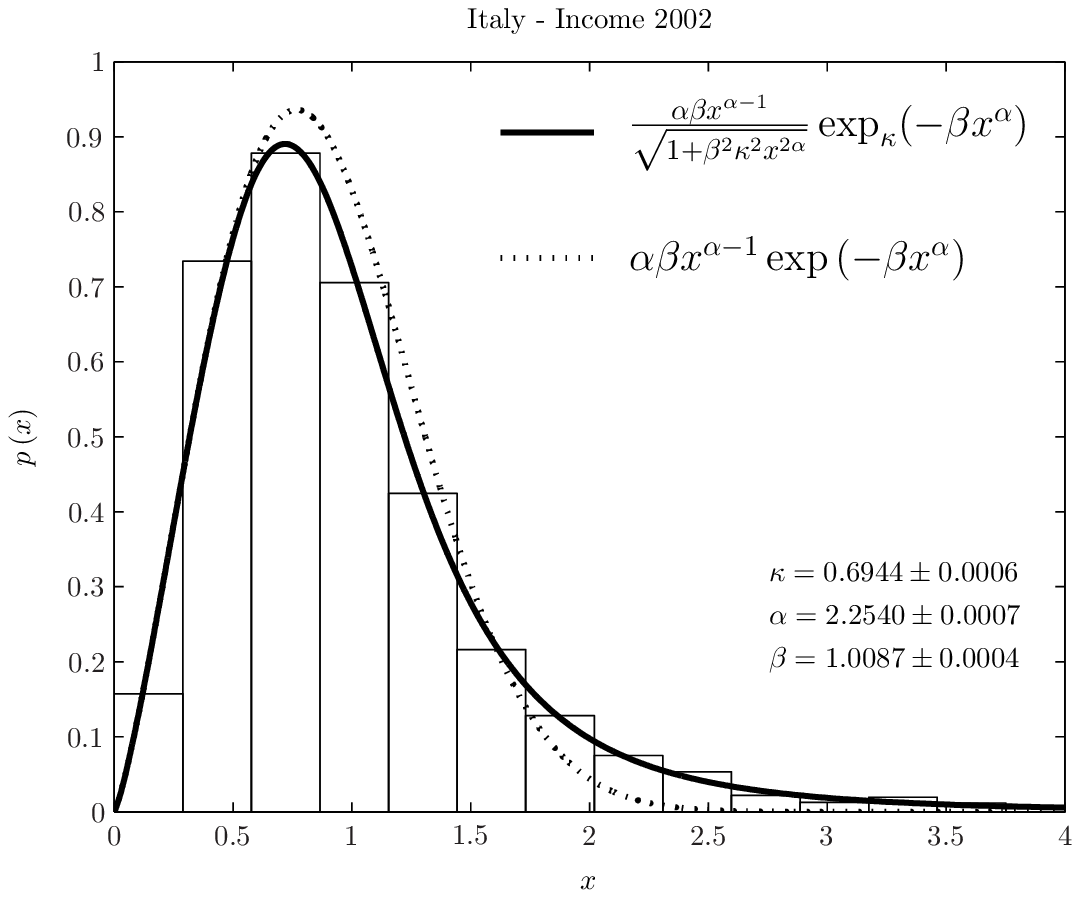}}}
\caption{Same plots as in FIG. \ref{fig:Figure4} for the Italian personal income distribution from the 2002 SHIW data file with $\kappa=0.6944\pm0.0006$, $\alpha=2.2540\pm0.0007$, and $\beta=1.0087\pm0.0004$. The income variable is measured in current year euros}
\label{fig:Figure5}
\end{center}
\end{figure}
\begin{figure}[t]
\centering
\begin{center}
\mbox{
\subfigure[\hspace{0.2cm}Complementary CDF]{\label{fig:Figure6_a}\includegraphics[width=0.48\textwidth]{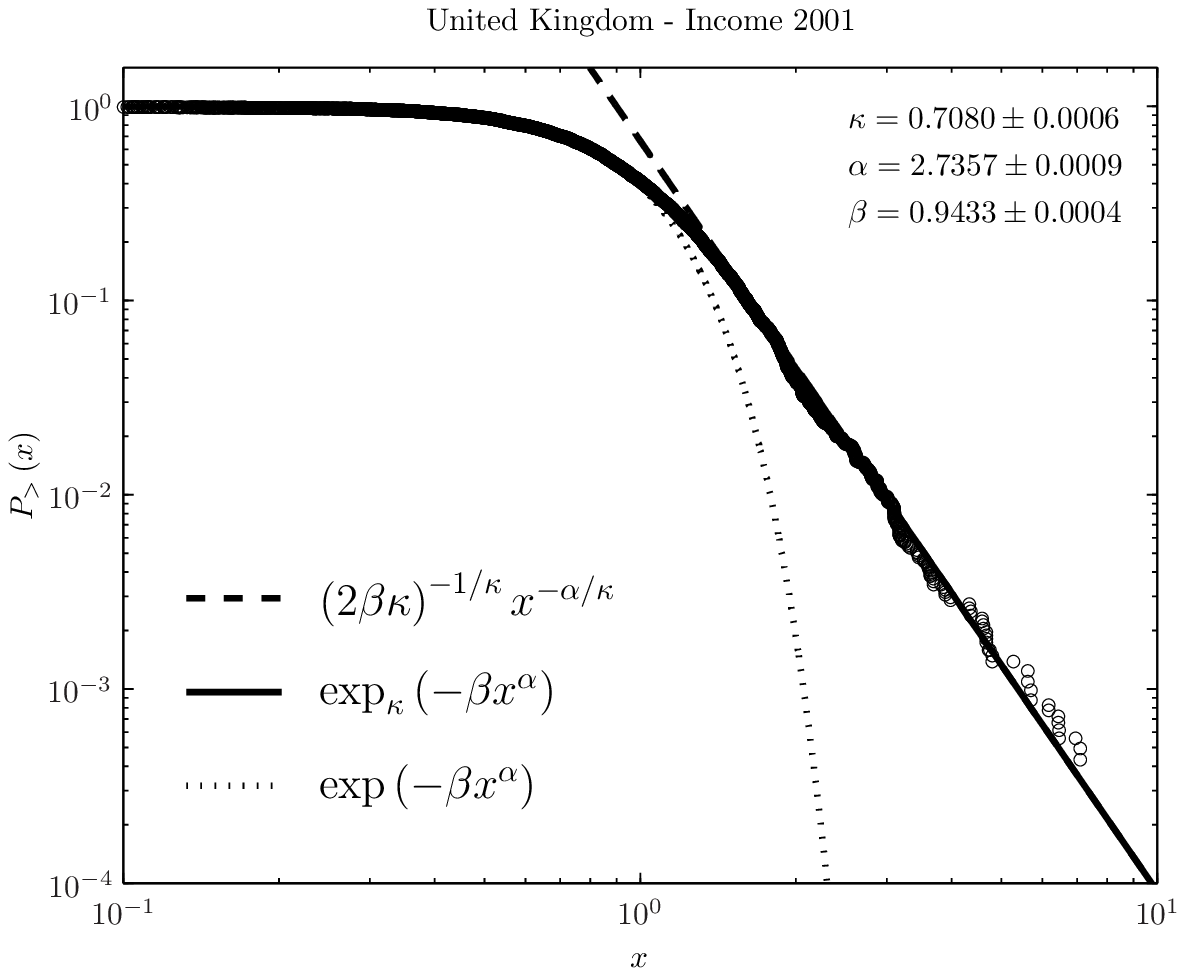}}
\subfigure[\hspace{0.2cm}PDF histogram plot]{\label{fig:Figure6_b}\includegraphics[width=0.48\textwidth]{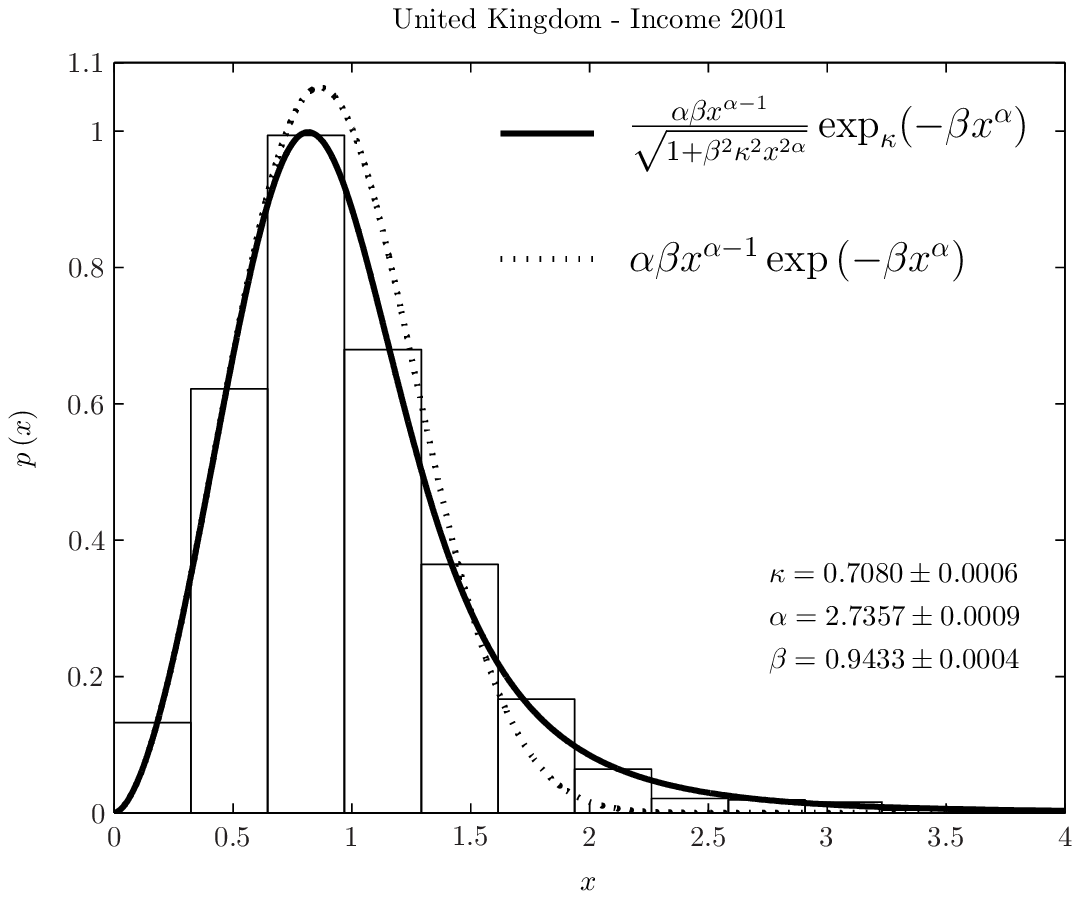}}}
\caption{Same plots as in FIGS. \ref{fig:Figure4} and \ref{fig:Figure5} for the UK personal income distribution from the 2001 BHPS-CNEF data file with $\kappa=0.7080\pm0.0006$, $\alpha=2.7357\pm0.0009$, and $\beta=0.9433\pm0.0004$. The income variable is measured in current year British pounds}
\label{fig:Figure6}
\end{center}
\end{figure}

The results obtained by fitting our theoretical model through the observational data are reported in TABLE \ref{tab:Table1} and FIGS. \ref{fig:Figure4}, \ref{fig:Figure5}, and \ref{fig:Figure6}.\footnote{To find the parameter values that give the most desirable fit, we have used the Constrained Maximum Likelihood (CML) estimation method \cite{Schoenberg1997}, which solves the weighted maximum log-likelihood problem
\[l\left(\mathbf{x};\theta\right)=\sum\limits^{n}_{i=1}\log p\left(x_{i};\theta\right)^{w_{i}},\]
where $n$ is the number of observations, $w_{i}$ is the survey weight accommodating features of the sample design and the population structure \cite{TheCanberraGroup2001}, $p\left(x_{i};\theta\right)$ is the probability of $x_{i}$ given $\theta$, the vector of parameters, subject to the non-linear equality constraint given by Eq. \eqref{eq:Equation17} and bounds $\alpha,\beta>0$ and $\kappa\in\left[0,1\right)$. The CML procedure finds values for the parameters in $\theta$ such that $l\left(\mathbf{x};\theta\right)$ is maximized using the Sequential Quadratic Programming (SQP) method \cite{Han1977} as implemented in \textsc{Matlab}\textsuperscript{\textregistered} 7. We have then calculated the approximate 95\% confidence interval half-width around each parameter by using the normal approximation
\[\theta\pm z_{1-\frac{\alpha}{2}}\cdot\sigma_{\theta},\]
where $\sigma_{\theta}$ denotes the estimate standard error\textemdash obtained from a finite difference approximation to the asymptotic covariance matrix of the maximum likelihood estimators of the parameters, and $z_{1-\frac{\alpha}{2}}$ is defined such that $\Phi\left(z_{1-\frac{\alpha}{2}}\right)=1-\frac{\alpha}{2}$, being $\Phi\left(\cdot\right)$ the standard normal distribution function. The overall analysis uses a simple equivalence scale adjusting income by the square root of the number of household members to account for differences in household size and composition.} Panel (a) of the figures shows the empirical cumulative distribution estimate\footnote{The empirical cumulative distribution is equal to the normalized sum of the survey weights of the individuals with incomes above $x$.} of $x$ along with three different curves in the log-log scale: the $\kappa$-generalized distribution, Eq. \eqref{eq:Equation10}; the ordinary stretched exponential (Weibull) distribution, Eq. \eqref{eq:Equation11}; the pure power-law distribution, Eq. \eqref{eq:Equation12}. In panel (b), the histogram of the reconstructed probability density\footnote{In order to estimate the empirical probability density, we divide the income axis into bins of width $\Delta x$, calculate the sum of the survey weights of the individuals with incomes from $x$ to $x+\Delta x$, and plot the obtained histogram.} is contrasted to the theoretical curves corresponding to Eqs. \eqref{eq:Equation13} and \eqref{eq:Equation14} with the same parameter values as in TABLE \ref{tab:Table1} and panel (a) of FIGS. \ref{fig:Figure4}, \ref{fig:Figure5}, and \ref{fig:Figure6}. It is clear that the $\kappa$-generalized distribution offers a great potential for describing the data over their whole range, from the low to medium income region through to the high income Pareto power-law regime, including the intermediate region for which a clear deviation exists when two different curves are used.\footnote{Pareto's contribution \cite{Pareto1896,Pareto1897a,Pareto1897b} has also stimulated further research on the specification of new models to fit the whole range of income\textemdash the interested reader is referred to the review in Ref. \cite{KleiberKotz2003} and the bibliography therein for an exhaustive list of personal income distributions and their basic properties. Weibull, gamma, beta, Dagum, Singh-Maddala, Fisk, Lomax, Pareto-L\'evy, Champernowne\textemdash just to name a few distributions many of which are special or limiting cases of more general parametric families, such as the generalized gamma distribution and the (generalized) beta distribution of the second kind\textemdash have all been used as descriptive models for the overall distribution of income. Although we are well aware of the existence of this numerous body of income distributions for which our work could ultimately result in duplication of effort, our main goal in this field is to concentrate on the opportunity of transposing the tools, methods and concepts from statistical mechanics to economics.}
\begin{table}[t]
\centering
\caption{Estimated parameters (with 95\% confidence intervals half-widths) of the $\kappa$-generalized distribution for the countries and years shown in FIGS. \ref{fig:Figure4}, \ref{fig:Figure5}, and \ref{fig:Figure6}. Also shown is the estimated weighted average income}
\label{tab:Table1}
\begin{ruledtabular}
\begin{tabular}{cccc}
\hline
&Germany&Italy&United Kingdom\\
\hline
$\kappa$&$0.5697\pm0.0005$&$0.6944\pm0.0006$&$0.7080\pm0.0006$\\
\hline
$\alpha$&$2.5659\pm0.0007$&$2.2540\pm0.0007$&$2.7357\pm0.0009$\\
\hline
$\beta$&$0.8788\pm0.0003$&$1.0087\pm0.0004$&$0.9433\pm0.0004$\\
\hline
$\left\langle z\right\rangle$&$36315.67\pm339.24$&$18087.92\pm246.85$&$14982.20\pm183.09$\\
\hline
\end{tabular}
\end{ruledtabular}
\end{table}


\section{Final remarks}
\label{sec:SectionIV}
Since the early study of Pareto, numerous recent empirical works have all shown that the power-law tail is an ubiquitous feature of income distributions. However, even 100 years after Pareto's observation, the understanding of the shape of income distribution is still far to be complete and definitive. This reflects the fact that there are two distributions, one for the rich, following the Pareto's law, and one for the vast majority of people, which appears to be governed by a completely different law.
\par
In the present work we have affirmed support for a new fitting function, having its roots in the framework of $\kappa$-generalized statistical mechanics, which shows to be able to describe the data over the entire range, including even the power-law tail. This distribution has a bulk very close to the stretched exponential one\textemdash which is recovered when the deformation parameter $\kappa$ approaches to zero\textemdash while its tail decays following a power-law for high values of income, thus providing a kind of compromise between the two description.
\par
The good concordance of our generalized statistical distribution with observational data on personal income may suggest a new path for investigating economic relations, namely the development of models within the framework of $\kappa$-generalized statistical mechanics.


\end{document}